\documentclass[preprint2]{aastex}

\slugcomment{tentatively scheduled for the v578 n1 pt1 ApJ October 10,
2002 issue}

\shorttitle{3C\,129 Jet'}
\shortauthors{Harris, Krawczynski, \& Taylor}

\begin{document}


\title{X-ray Detection of the Inner Jet in the Radio Galaxy 3C\,129}


\author{D. E. Harris}
\affil{Smithsonian Astrophysical Observatory, 60 Garden Street,
Cambridge, MA 02138}
\email{harris@cfa.harvard.edu}

\author{H. Krawczynski}
\affil{Yale University, P.O. Box 208101, New Haven, CT 06520-8101}
\email{krawcz@astro.yale.edu}

\and

\author{G. B. Taylor}
\affil{NRAO, Box O, Socorro NM 87801}
\email{gtaylor@cv3.cv.nrao.edu}



\begin{abstract}
During the course of an investigation on the interaction of the radio
galaxy 3C\,129 and its ambient cluster gas, we found excess X-ray
emission aligned with the northern radio jet.  The emission extends
from the weak X-ray core of the host galaxy
$\approx~2.5^{\prime\prime}$ to the first resolved radio knot.  On a
smaller scale, we have also detected a weak radio extension in the
same position angle with the VLBA.  Although all the evidence suggests
that Doppler favoritism augments the emission of the northern jet, it
is unlikely that the excess X-ray emission is produced by inverse
Compton emission.  We find many similarities between the 3C\,129 X-ray
jet and recent jet detections from Chandra data of low luminosity
radio galaxies.  For most of these current detections synchrotron
emission is the favored explanation for the observed X-rays.

\end{abstract}




\section{Introduction}

The radio galaxy 3C\,129 is a low luminosity (FRI type) 'tailed radio
galaxy' seen in projection towards the outer edge of the X-ray
emission from the hot gas of a nearby cluster of galaxies (Leahy and
Yin, 2000, Taylor et al. 2001).  Since the cluster lies at low
galactic latitude towards the anti-center, it has not been well
studied in the optical.

We obtained Chandra observations in order to study the interaction of
the radio structures with the hot intra cluster medium (ICM) and that
work will be presented elsewhere (an analysis of the ICM properties
has been performed by Krawczynski 2002, and a paper on pressure
balance is in preparation).  In this paper we report on faint X-ray
emission detected from the core of the 3C\,129 galaxy and from the inner
3 kpc of the northern radio jet.  We include the results of
'follow-up' observations with the VLBA \footnote{The National Radio
Astronomy Observatory is operated by Associated Universities, Inc.,
under contract with the National Science Foundation.} in
sec.~\ref{sec:radio}.

X-ray emission from radio jets presents us with the problem of
identifying the emission process but once this process is determined,
we can then obtain new constraints on physical parameters (Harris and
Krawczynski, 2002).  With the introduction of the relativistic beaming
model of Celotti (Celotti, Ghisellini, \& Chiaberge, 2001) and
Tavecchio (Tavecchio, et al. 2000), most X-ray emission from jets has
been interpreted as indicating either synchrotron emission or inverse
Compton scattering off the cosmic microwave background (CMB).  For
3C\,129, we show that synchrotron emission is the probable process, as
has been found for a number of other FRI radio galaxies (Worrall,
Birkinshaw, \& Hardcastle, 2001; Hardcastle, Birkinshaw, \& Worral
2001).  The implications of the detected X-ray emission are discussed
in sec.~\ref{sec:disc}.

The redshift of the radio galaxy at the center of the cluster, 3C\,129.1
is z=0.0208 (Spinrad, 1975) and we take this for our distance estimate
of D$_L$=126~Mpc with H$_o$~=~50~km~s$^{-1}$~Mpc$^{-1}$ and q$_o$~=~0.
One arcsec then corresponds to 0.60 kpc.

\section{X-ray data}\label{sec:xray}

The X-ray observation was obtained with the ACIS-S detector on the
Chandra Observatory (obsid 2218, 2000Dec09).  The exposure time was
31.46 ksec and the 3C\,129 galaxy was observed with the back
illuminated `S3' chip.  After standard Chandra pipeline processing
(R4CU5UPD12.1 on 2000Dec12) we rejected intervals with excess counting
rates (indicative of particle flares) resulting in a livetime of
30.405 ksec.  Events with energies less than 0.3keV or greater than 8
keV were rejected.

\begin{figure}
\plotone{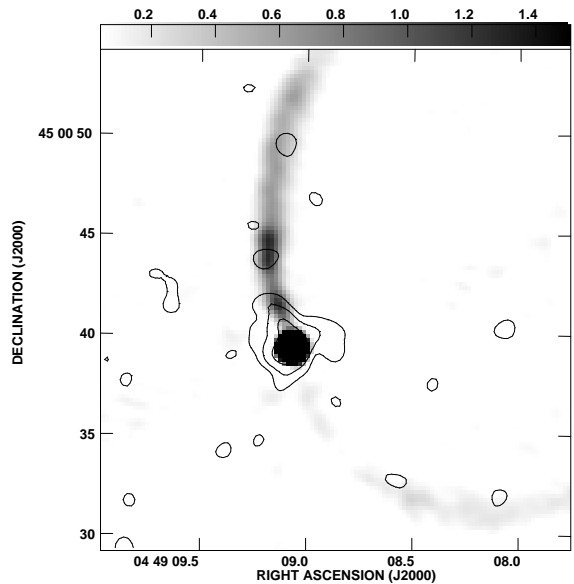}
\caption{An 8 GHz VLA map (greyscale) of the inner part of 3C\,129
with X-ray contours overlayed.  The grey scale is from 0.05 to 1.5
mJy/beam and the beamwidth is 0.83$^{\prime\prime}$ FWHM.  The X-ray
data were smoothed with a Gaussian of FWHM=1$^{\prime\prime}$;
contours are logarithmic, increasing by factors of 2 with the lowest
contour at 1.8 counts per square arcsec (2.27 times the background
level).  The X-ray image has been shifted 0.24$^{\prime\prime}$ in RA
to align the radio and X-ray cores. \label{fig:over}}
\end{figure}

We then binned the data by a factor of 1/4 to obtain images with pixel
size 0.123$^{\prime\prime}$.  Various Gaussian smoothing functions
were then convolved with the data and one example is shown in
figure~\ref{fig:over}, an overlay of the radio image with X-ray
contours.  While it is clear that there is excess X-ray emission
coincident with the first visible radio knot, `N2.3', it appears that
the X-ray morphology is essentially a projection from the core rather
than a completely resolved separate structure.  There is also a 1 to 2
$\sigma$ excess located at the beginning of the second radio knot,
`N5.0'.  All of these features are weak.  For a circular aperture of
radius 0.9$^{\prime\prime}$, we find only 30 net counts in the core
and an additional 12 net counts defining the jet.  N5.0 contains only
4 net counts.

The observation was performed with a stage offset ('sim z') of -5.86mm
(119.5$^{\prime\prime}$ or 243 pixels toward the readout edge), a y
offset of -1$^{\prime}$ to move the target to the center of a node, and a
specified roll angle so as to position the 0.4$^{\circ}$ radio tail on
the ACIS-S array.  Since the target position was not the
center of the galaxy, this procedure resulted in the core of the 3C\,129
galaxy being 90$^{\prime\prime}$ from the optical axis.

To check on the reality of the jet morphology, a 1.49keV point spread
function (PSF) was generated to match the location off-axis and the
pixel size of 0.123$^{\prime\prime}$.  This PSF image was then
smoothed with a 1$^{\prime\prime}$ Gaussian.  The resulting image has
quasi circular contours with radius of 0.66$^{\prime\prime}$ for the
50\% intensity levels.  This value can be compared with
0.8$^{\prime\prime}$ for 3C\,129 in directions to the south and
south-west (away from the jet) and 1.2$^{\prime\prime}$ for the 50\%
contour in the position angle of the jet.  If the X-ray jet were to be
caused by statistical happenstance, it's alignment with the radio jet
would be coincidental.

\begin{figure}
\plotone{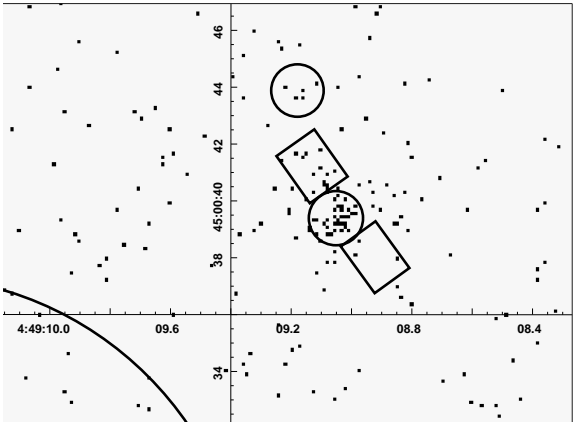}
\caption{A map of the X-ray photons with the apertures chosen for
intensity measurements overlayed.  The segment of the large circle to
the lower left is part of the circle used for background estimates.
The circle with the obvious concentration of counts has a radius of
0.95$^{\prime\prime}$ and serves to measure the core.  The abutting
rectangle to the NE is for the X-ray jet extending out to the N2.3
radio knot and the small circle beyond corresponds to the first part
of the radio knot N5.0.  The rotated box to the SW is a control region
for the jet emission.  We thank W. Joye for his work on the imaging
tool, ds9 which allows us to make high quality figures with minimal
effort.  \label{fig:regions}}
\end{figure}

To asses the various emission mechanisms for the X-rays, we need to
define areas (and their implied emitting volumes) and measure
intensities.  These regions were selected on the basis of the smoothed
map (fig.~\ref{fig:over}), but the measurements were made on the event
file.  Mindful of the paucity of X-ray photons, we are content
with order of magnitude estimates.  For the core, we have taken a
circle of radius 0.95$^{\prime\prime}$; for the X-ray jet we use a
rotated box of dimensions
2.03$^{\prime\prime}~\times~1.63^{\prime\prime}$; and for the N5.0
feature we use a small circle of radius 0.92$^{\prime\prime}$.  These
regions are shown in fig.~\ref{fig:regions}.

Using the PIMMS tool and XSPEC/fakeit with a power law
spectrum, we find a conversion value for 1~c/s (0.3 to 8 keV) to
unabsorbed flux, f$_x$(0.5-5keV) of 1.11 ($\alpha$=0.5); 1.16
($\alpha$=1.0); and 1.27 ($\alpha$=1.5)
$\times~10^{-11}$~erg~cm$^{-2}$~s$^{-1}$.  This allows us to determine
rough fluxes for the features measured.

\section{Radio data}\label{sec:radio}

The VLA data used in this paper are those described in Taylor et
al. (2001).   However, we mainly used the 8 GHz data at their inherent
resolution of 0.83$^{\prime\prime}$ FWHM rather than the versions 
previously published which were smoothed to larger beams so as to
match lower frequency data.  This beam size is quite close to what
we obtained with Chandra so meets the need of obtaining comparative
morphologies and corresponding flux densities.

To obtain some sense of what role Doppler boosting might play near the
radio core of 3C\,129, on 16 December 2001 we observed the core at
4.986 GHz with the 10 element VLBA. Because of inclement weather, no
data were obtained from the VLBA antenna at Mauna Kea.  A total
bandwidth of 32 MHz was recorded in left circular polarization only
using 2 bit sampling.  The VLBA correlator produced 16 frequency
channels across each 8 MHz wide IF during every 2 second integration.
Amplitude calibration for each antenna was derived from measurements
of the antenna gain and system temperatures during each run.  Delays,
rates and phases were derived from the nearby (2.79 degrees distant)
calibrator J0440+4244 and transferred to 3C\,129.  A 3 minute cycle of
120s:60s on target:calibrator was used.  To check the quality of the
phase referencing the calibrator J0427+4133 was observed 5 times
during the 4 hour run.  The coherence on J0427+4133 (2.58 degrees
distant from J0440+4244) was found to be $\sim$85\%.

\begin{figure}
\plotone{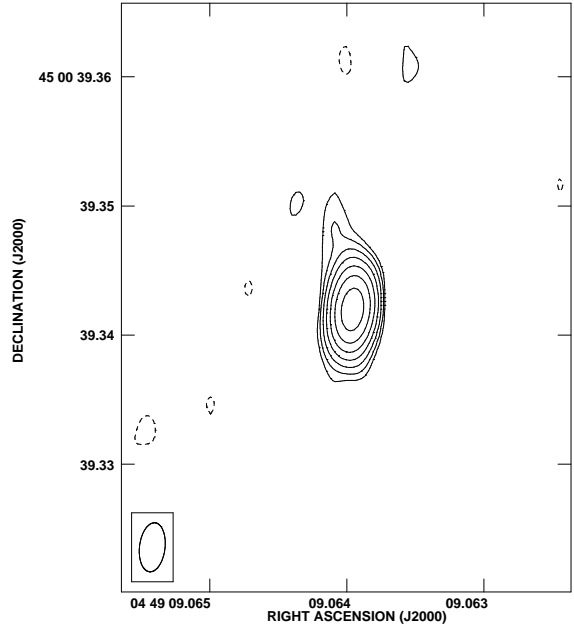}
\caption{The VLBA 5GHz map.  The beamwidth is
3.8~mas~$\times$~2.0~mas in PA=-8$^{\circ}$ and
the contours are logarithmic, increasing by factors of 2 in
brightness.  The lowest contour is 0.25~mJy/beam and the peak intensity
is 26.9 mJy/beam.\label{fig:vlba}}
\end{figure}

Once delay and rate solutions were applied the data were averaged in
frequency over 32 MHz.  The data from all sources were edited and
averaged over 20 second intervals using DIFMAP (Shepherd,
Pearson \& Taylor 1995) and then were subsequently self-calibrated and
imaged.  The final image is shown in Fig.~\ref{fig:vlba}.  The
position of the core of 3C\,129 derived from modelfitting with 
DIFMAP is (J2000) RA 04h49m9.06396 DEC +45d00'39.342.  Based on the
observed offset of J0427+4133, the accuracy of this position should be
$\sim$0.35 mas.  The peak flux density is $\approx$~2/3 that obtained
some years earlier with a 1.8$^{\prime\prime}$ beam (Taylor et
al. 2001).

The detection of the northern jet extending 6pc from the core supports
the notion that the northern jet is the one coming towards us (in
agreement with the VLA morphology) and that Doppler favoritism is
operating on the pc scale.  The PA of the 6pc scale feature is
13$^{\circ}$; essentially the same as the value measured 630 pc from
the core on the VLA map (PA~$\approx~14^{\circ}$).  Thus we may expect
very little bending in the jet up to about 2kpc (3.4$^{\prime\prime}$)
and this inner straight segment of the jet is the part that is
detected by Chandra.

\section{Parameters for Emission Models}

To estimate physical parameters associated with various X-ray emission
mechanisms, we need to assume values for some unmeasurable parameters
such as the spectral index and refine volume estimates.  We will also
need to estimate the radio flux densities which correspond to the
X-ray emitting volumes, not to the obvious radio features.  For the
radio spectral index we use $\alpha~=~0.8$.

For the jet, we take a cylinder of length 1.8$^{\prime\prime}$ and radius
0.25$^{\prime\prime}$.  For this volume, we ascribe a flux density of
3mJy at 8 GHz and f$_x$(0.5-5)= $\frac{12}{30403}~\times~1.16~10^{-11}
= 4.6~10^{-15}$~erg~cm$^{-2}$~s$^{-1}$.  Since the X-ray emission is
brighter towards the core end of this cylinder, and the radio is
brighter at the down-stream end, this is a gross approximation.


\subsection{Thermal Bremsstrahlung Emission}

The log of the X-ray luminosity (0.5-5keV) for the jet is 39.967
(erg~s$^{-1}$) and the density required to produce this emission from
the cylindrical volume would be 0.7cm$^{-3}$.  The mass of the
emitting cylinder would be 1.4~$\times~10^6~M_{\odot}$.  Assuming a
temperature of 2 keV means that the pressure would be
4.4$~\times~10^{-9}$dyne~cm$^{-2}$.  This pressure can be compared to
that of the ICM at this location:
0.0075~$\times~10^{-9}$dyne~cm$^{-2}$ (Krawczynski, 2002).  Since this
volume is well inside the galaxy, there could be additional pressure
contributed by cooler gas which does not produce X-rays.

\subsection{Synchrotron Emission}

At similar resolutions, the X-ray emission decreases monotonically
moving away from the core, whereas the radio brightness is low
adjacent to the unresolved (0.83$^{\prime\prime}$ beam) core and then
increases to the enhancement we call 'knot N2.3'.  Because of this
discrepancy in morphology, a simple synchrotron model is not
easily constructed.

If the morphology difference arises by a statistical fluctuation from
the small number of photons defining the X-ray jet, we may calculate
the synchrotron parameters necessary to produce the observed X-rays.
For the radio emission from N2.3 (10$^7$ to 10$^{11}$~Hz), the log of
the luminosity would be 39.517 erg~s$^{-1}$ and the equipartition
field would be 40 $\mu$G. For a synchrotron X-ray model, we need to
extend the radio spectrum up to 10$^{18}$~Hz with the spectral index
$\alpha$=0.9 (c.f. the radio spectral index between 5 and 8 GHz is in
the range 0.55 to 0.85 with a beamsize of 1.8$^{\prime\prime}$, Taylor
et al. 2001).  In this case the log of the luminosity would be 40.669
erg~s$^{-1}$ and the equipartition field would be 42 $\mu$G.  Although
there would be only an insignificant change in the total energy
contained in the source, the power law distribution would have to
extend to $\gamma=7~\times~10^7$ with a halflife of some 60 years for
electrons of this energy.

If the morphology difference is real, then the radio flux density to
associate with the X-ray jet will be a factor of two or three less
than the 3mJy found for knot N2.3.  However, that would change the
values derived above very little and the only different ingredient
would be the natural picture of a quite limited region of shock
acceleration capable of producing the high energies required for
X-rays, followed by a further (downstream) segment of the jet where
acceleration of the more common energies continues.

\subsection{Inverse Compton Emission}

The synchrotron self-Compton model fails because the photon energy
density is so low that the predicted flux would be 4 orders of
magnitude below that observed (assuming an equipartition field of
40~$\mu$G).

IC scattering off the CMB photons would require a magnetic field
strength of 0.3~$\mu$G, more than a factor of 100 below the
equipartition field and the emitting volumes appear not to
coincide as expected from IC emission.

Even if we invoke relativistic beaming and ignore the disparity in
morphology between radio and X-rays, to produce the observed X-ray jet
would require an angle between the jet velocity vector and the line of sight
of 7$^{\circ}$ or less and a beaming factor of 8.  If there is a
difference in morphologies, the beaming parameters become more
stringent since the corresponding radio flux density to associate with
the X-ray emission would be less than used in the calculations.  

These values are inconsistent with estimates from the radio data.
From the observed ratio of intensities of the inner radio jets (4.46), the
angle between the line of sight and the N jet has to be less than
75$^{\circ}$ and is most likely greater than 30$^{\circ}$ since
we see the two sides of the jet nowhere near lying on top of each
other.  This range in angles corresponds to beaming factors in the
range 1.16 to 1.3 and jet fluid velocities, $\beta~(=~\frac{v}{c}$), in
the range 0.3 to 1.

\section{Discussion}\label{sec:disc}

Granted that we are dealing with few photons and thus an insecure
morphology, we believe the evidence favors synchrotron emission for
the observed X-rays.  Undoubtedly there are bulk relativistic
velocities in the jet producing the observed intensity differences
between the N and S jets, but with velocity vectors not too far from
the plane of the sky, we see only mild boosting and the parameters for
IC/CMB emission are completely at odds with all other evidence.

If the bulk of the detected X-ray emission is in fact upstream of the
radio knot N2.3, it would simply indicate that an acceleration region
capable of producing $\gamma$ of order 10$^7$ to 10$^8$ would be
followed by a more extensive acceleration region incapable of such
high energies, but rather producing up to $\gamma~\approx~10^4$.  Even
in weak fields of order 30 to 50 $\mu$G, the half-life for electrons
producing X-rays is so short that they could travel no more than 30pc
from their acceleration region.  Thus the X-rays clearly demarcate
that sort of acceleration region.

There are now several detections of X-ray emission from jets in FRI
radio galaxies.  For M87 (Marshall et al. 2002) the radio, optical,
and X-ray morphologies are quite close if not identical but upstream
offsets of X-ray brightness peaks compared to those of the radio have
been documented for 3C\,66B (Hardcastle et al. 2001) and 3C\,31
(Hardcastle et al. 2002).  For these sources the offsets are a few
hundred pc, slightly smaller than our (uncertain) value of 480pc for
3C\,129.  The situation in the jet of Cen A (Kraft et al. 2000) is
confused with some features aligning well at radio and X-ray bands,
but for others it is not always clear which features correspond at
the other wavelength.

In Table~\ref{tab:fr1} we give comparative values of size and
luminosities for several FRI detections.  It can be seen that the
3C\,129 parameters are quite consistent with the others.  While we
cannot rule out thermal bremsstrahlung as the cause of the X-rays from
3C\,129, it seems likely that as for the other FRI detections,
synchrotron emission is the favored process.



\acknowledgments

This work was partially supported by NASA grants and contracts. At
SAO: GO1-2135A and NAS8-39073; and at Yale: GO 0-1169X.



References\\



\noindent
Celotti, A., Ghisellini, G., \& Chiaberge, M. 2001 \mnras\ 321, L1-5

\noindent
Hardcastle, M.J., Birkinshaw, M., \& Worral, D.M. 2001 \mnras\ 326, 1499

\noindent
Hardcastle, M.J., Worral, D.M., Birkinshaw, M., Laing, R.A. \& Bridle,
A.H. 2002, \mnras\ (in press)

\noindent
Harris,  D. E. and Krawczynski, H. 2002 \apj\ 565, 244

\noindent


\noindent
Kraft, R.P. et al. 2000 \apj\ 531, L9

\noindent
Krawczynski, H. 2002, \apj\ (in press)

\noindent
Leahy, D. A. and Yin, D. 2000, \mnras\ 313, 617

\noindent
Marshall, H.L., Miller, B.P., Davis, D.S., Perlman, E.S., Wise, M.,
Canizares, C.R., and Harris, D.E. 2002, ApJ 564, 683

\noindent
Shepherd, M.~C., Pearson, T.~J., \& Taylor,
G.~B. 1995, BAAS, 27, 903

\noindent
Spinrad, H. 1975, \apj\ 199, L1

\noindent
Tavecchio, F., Maraschi, L., Sambruna, R.M., \& Urry, C.M. 2000 \apj\
544, 23

\noindent
Taylor, G.B., Govoni, F., Allen, S.A., \& Fabian, A.C. 2001 \mnras\
326, 2

\noindent
Worrall, D.M., Birkinshaw, M., \& Hardcastle, M.J. 2001, \mnras\ 326, L7



\onecolumn


\begin{deluxetable}{lcccl}
\tablecaption{Parameters for X-ray jets in FRI radio galaxies. \label{tab:fr1}}
\tablewidth{0pt}
\tablehead{
\colhead{Source} & \colhead{Scale}   & \colhead{Projected Length}   &
\colhead{L$_x$(0.5-5keV)} &
\colhead{Reference}\\
 & \colhead{kpc/arcsec} & \colhead{(pc)} &
\colhead{10$^{40}$erg~s$^{-1}$}     & \\
}
\startdata
3C129 & 0.60 & 1320 & 0.9 & this paper \\
3C31  & 0.48 & 3356 & 4.9 & Hardcastle et al. 2002 \\
B2 0206+35 & 1.02 & 2040 & 16.0 & Worrall et al. 2001\\
3C66B & 0.61 & 4270\tablenotemark{a} & 13.0 & Hardcastle et al. 2001\\
B2 0755+37 & 1.17 & 2574 & 39.0 & Worrall et al. 2001\\
M87 & 0.08 & 1386 & 7.8 & Marshall et al. 2002\\
Cen A & 0.02 & 3400 & 0.3 & Kraft et al. 2000\\
 \enddata


\tablenotetext{a}{Most of the X-ray intensity is closer than 1220 pc
from the core.}



\end{deluxetable}

\end{document}